# Network Simulation with Complex Cyber-attack Scenarios

Tiago Dias[0000-0002-1693-7872], João Vitorino[0000-0002-4968-3653], Eva Maia[0000-0002-8075-531X] and Isabel Praça[0000-0002-2519-9859]

Research Group on Intelligent Engineering and Computing for Advanced Innovation and Development (GECAD), School of Engineering, Polytechnic of Porto (ISEP/IPP), 4249-015 Porto, Portugal
{tiada,jpmvo,egm,icp}@isep.ipp.pt

**Abstract.** Network Intrusion Detection (NID) systems can benefit from Machine Learning (ML) models to detect complex cyber-attacks. However, to train them with a great amount of high-quality data, it is necessary to perform reliable simulations of multiple interacting machines. This paper presents a network simulation solution for the creation of NID datasets with complex attack scenarios. This solution was integrated in the Airbus CyberRange platform to benefit from its simulation capabilities of generating benign and malicious traffic patterns that represent realistic cyber-attacks targeting a computer network. A realistic vulnerable network topology was configured in the CyberRange and three different attack scenarios were implemented: Man-in-the-Middle (MitM), Denial-of-Service (DoS), and Brute-Force (BF).

## 1 Introduction

Traditional Network Intrusion Detection (NID) systems can be bypassed by malicious attackers, since these systems are usually built on top of well-known rules and need extensive analysis by experts in the field [1]. As such, there is a need to find more suitable and adaptable solutions. Machine Learning (ML) models have demonstrated promising results in the NID field by learning complex network features and being able to detect complex cyber-attacks and even zero-day attacks [2]. However, these models require a great amount of high-quality data to be able to generalize well, which is difficult to obtain from a real computer network with missing or corrupted information [3]. Therefore, simulating multiple interacting machines could be an interesting approach to capture reliable traffic of a computer network by simulating machines, servers, live traffic and cyber-attacks.

This paper presents a network simulation solution for the creation of NID datasets with complex attack scenarios. This solution was integrated in the Airbus CyberRange platform to benefit from its simulation capabilities of generating benign and malicious traffic patterns that represent realistic cyber-attacks targeting a computer network. A realistic vulnerable network topology was configured in the CyberRange and three different attack scenarios were implemented: Man-in-the-Middle (MitM), Denial-of-Service (DoS), and Brute-Force (BF).

This paper is organized into multiple sections. Section 2 provides a survey of previous works on the use of network simulation platforms for network traffic flow generation. Section 3 presents the proposed solution for the simulation of more complex attack scenarios, which will be used to generate a new dataset in future work. Section 4 presents the main conclusions and the outlined future work.

## 2   State-of-the-Art

The simulation of computer networks has been a focal point of research for several tasks including cybersecurity, which lead to the creation of several simulation platforms. Therefore, it is important to survey the most significant approaches and their extensions, comparing their benefits, limitations, and contributions to the field.

Built on top of the discrete network simulator NS3, which supports the simulation of various protocols, routing algorithms, and wired and wireless network types, Chengze et al. [4] proposed NS4, adding P4 programming language support to improve the data plane in software-defined networks and simplify the programming of switches. GNS3 [5] is an open-source tool that focuses on the simulation of the networks physical layer, being capable of emulating licensed hardware devices. It allows simulated interactions between simulated networks and real interactions with real network. It also allows for traffic monitoring by permitting the extraction of PCAPs. OMNeT++ [6] is yet another network simulator that can be combined with INET framework, a network traffic generator.

Aside from open-source simulators, NetSim [7] is also a highly configurable discrete-event network simulator that allows the simulation of network hosts and traffic that integrates network analysis tools such as Wireshark and Matlab. Cisco Packet Tracer [8] offers a proprietary network simulator, for educational purposes of the Cisco Networking Academy. It offers a user interface to build, configure, and experiment with networks and corresponding protocols, network types, and technologies.

Recently, Meyer-Berg et al. [9] have proposed a multi-agent simulation tool to specifically address the lack of network data in the NID field in the context of an IoT network. The tool is composed of five layers: Smart Devices, Ports, Protocols, Connections, and Links. Smart Devices act as agents, simulating sensors and actuators with internal states and multiple ports for data generation and reception. Ports are connection endpoints with states like CLOSED, OPEN, and SENDING, featuring configurable message intervals and delays. Protocols manage application-layer communications. Connections encapsulate protocol packets and may add metadata such as TCP Syn/Ack packets. Links model lower OSI layers and physical connections, defining packet loss and transmission delays.

The authors also implemented algorithms for generating small perturbations in the data to imitate a more realistic approach. They defined a scenario in a smart-home setting, featuring several IoT devices, manageable via a smart hub. In their scenario they executed DoS and DDoS attacks and were able to capture data to train a NID ML model. Their results show that K-Means algorithm was successful in detecting the cyber-attacks. However, the authors only provide the confusion matrix metrics and the time elapsed for train and detection, and so it is not clear if the algorithm generalizes well or is well fit to the data simulated.

Nevertheless, more recently, Airbus has released a novel licensed multi-purpose CyberRange [10], which is a platform that enables the simulation of many different network hosts and benign and malicious traffic via the configuration of many different scenarios. The scenarios automate the actions of the agents and permit the execution of custom commands in each host. Furthermore, since each host has its own OS running on top of VMWare, all traffic and analysis can be performed seamlessly.

Therefore, this paper explores this simulation platform that enables the customization of the actions of the network hosts in a scenario-based approach, facilitating traffic generation for NID dataset generation.

## 3  Proposed Solution

To produce NID datasets containing recent cyber-attacks in more complex attack scenarios, the Airbus CyberRange platform was utilized and improved to simulate network patterns of multiple interacting machines. It was required to configure a realistic vulnerable network topology that contains benign and malicious activity, and several types of network attacks. The following subsections describe the network topolog, and the considered attack scenarios with interactions between hosts.

### 3.1  Simulated Network Topology

The topology utilized in the proposed solution represents a relatively simple computer network that is common across small and medium enterprises. It is composed of three different subnets: (i) Service LAN, (ii) User LAN, and (iii) SOC LAN, which are interconnected through a central router (Fig. 1).

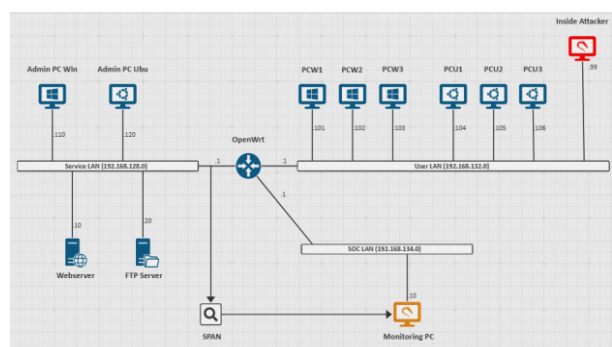

**Fig. 1.** Simulated Network Topology.

The Service LAN, located on the left of the router, is at the 192.168.128.0/24 subnet and contains all the business application services running, which includes an FTP

Server, a Webserver running an HTTP e-commerce store application, and two admin computers running Windows 10 and Ubuntu OS that are SSH enabled. In this setting, all network hosts are vulnerable to network attacks and generate live traffic of multiple different protocols, producing benign traffic.

The User LAN, located on the right of the router, is at the 192.168.132.0/24 subnet and simulates the activity of an enterprise user network. It includes six hosts running Windows 10 and Ubuntu OS, which continuously use the services running in the Service LAN. In this network there is also a Kali Linux, which acts as an attacker that was able to gain access to the internal network and intends to disrupt the backend services by executing different types of network attacks.

The SOC LAN, which is at the 192.168.134.0/24 subnet, acts as monitoring network and runs a Kali Linux host to monitor the traffic related to the Service LAN via the SPAN. A SPAN is a mechanism for mirroring the traffic traversing through the services network which includes the benign and malicious traffic.

### 3.2 Attacks Scenarios

According to the topology implemented, three automated attack scenarios targeting different services from the Service LAN were defined. These attack scenarios include MitM, DoS and BF, and define the possible interactions between the hosts of the topology. In general, all attacks are performed by the Kali Linux host in the User LAN targeting the Service LAN HTTP, SSH and FTP services and the remaining network hosts produce benign live traffic via the scenarios.

The MitM attack scenario highlights that the cyber-attack actions are executed in parallel with the live traffic generation ones during the simulation. The attacker starts by fingerprinting the network by executing a network scan to discover IPs, OS's, opened ports and services running. Afterwards, the attacker starts the ARP Poisoning attack. In parallel, multiple hosts from the User LAN communicate with Service LAN, generating HTTP and NTP data (Fig. 2).

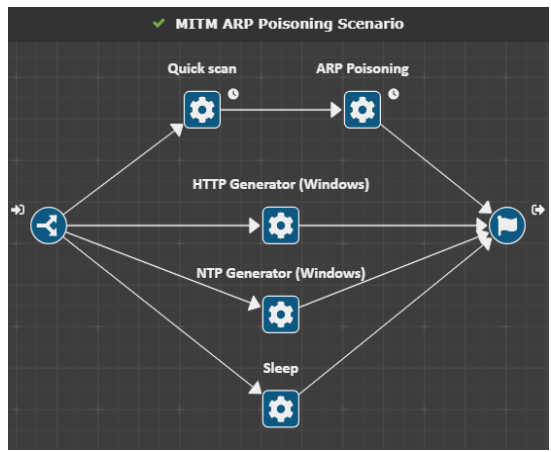

**Fig. 2.** Man-in-the-Middle Attack Scenario.

The DoS attack scenario is executed also in parallel, where the live traffic actions are autonomously executed by the benign hosts, and three flooding attacks, PUSH and ACK flood, ICMP and IGMP flood and TCP Connection Killer, are executed in the vulnerable machines running the services. The live traffic includes HTTP requests to the website, pings and ssh connection attempts to the ssh enabled hosts (Fig. 3).

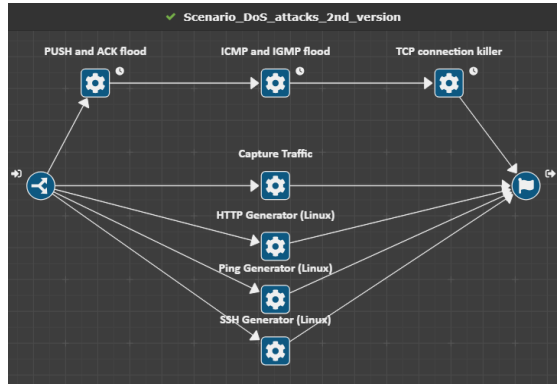

**Fig. 3.** Denial-of-Service Attack Scenario.

Similarly, the BF attack scenario is also executed by the topology hosts, where the live traffic actions are performed autonomously by the benign hosts and two BF attacks are executed by the Kali Linux host. Specifically in this scenario, the attacker is forced to sleep for 30 minutes between attacks to obfuscate the attack making it seem more realistic and increasing the difficulty to detect it (Fig. 4).

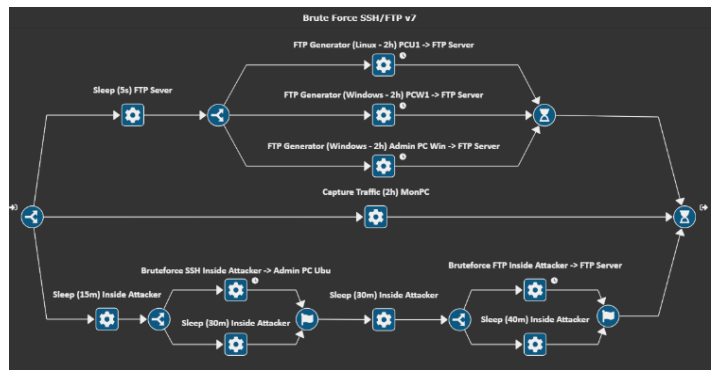

**Fig. 4.** Brute-Force Attack Scenario.

## 4 Conclusions and Future Work

This paper presented a network simulation solution implemented in the Airbus CyberaRange platform for the creation of NID datasets with complex attack scenarios. The scenarios configured included several benign activity actions to be executed by the network hosts, and malicious traffic patterns that represent realistic MitM, DoS, and BF attacks. In the future, the developed topology and the configured scenarios will be leveraged to generate more realistic NID datasets, which will be benchmarked to compare them to other publicly available datasets.

**Acknowledgements.** This work was supported by the CYDERCO project, which has received funding from the European Cybersecurity Competence Centre under grant agreement 101128052. This work has also received funding from UIDB/00760/2020.